\documentclass{ws-procs9x6}

\begin{document}

\title{The Work of Behram Kursunoglu}

\author{Philip~D. Mannheim\footnote{\uppercase{T}his work has
been supported in part by the \uppercase{U.~S.}
\uppercase{D}epartment of \uppercase{E}nergy
\uppercase{(D.O.E.)} grant No.
\uppercase{DE-FG02-92ER40716.00.}}}

\address{Department of Physics \\
University of Connecticut \\ 
Storrs, CT 06269, USA\\ 
E-mail: mannheim@uconnvm.uconn.edu\\}

\maketitle

\abstracts{Talk presented at the 2003 Coral Gables conference in honor and
appreciation of the work of Professor Behram Kursunoglu, general
relativist extraordinaire and founder of the Coral Gables series of
conferences, whose untimely death occurred shortly before the
2003 conference.}

Professor Behram Kursunoglu died shortly before the 2003 Coral Gables
conference, the latest in a long line of Coral Gables conferences which
Behram had been organizing since 1964. This particular conference had
originally been planned to honor Professor Paul Frampton on the occasion
of his sixtieth birthday, but with Behram's untimely passing the
conference additionally became a memorial to him, to his work and to his
life. 

At the conference Behram had been scheduled to speak on the topic "Will
Einsteinian theories continue to dominate twenty-first century physics?".
And even though I had not had an opportunity to discuss his talk with
him, I believe I know what he was going to talk about. Specifically, a few
years earlier at a previous Coral Gables conference at the turn of the
millennium, Behram had commented to me about the fact that Time magazine
had named Albert Einstein to be the man of the twentieth century. Behram
remarked to me that in another hundred years Time magazine would declare
Einstein to be the man of the twenty-first century as well. Behram felt
that up to now we had only just begun to skim the surface of the
implications of general relativity, and that during the upcoming century
general relativity would prove to be the foundation from which even
greater and deeper physics was going to arise. 

It should come as no surprise that Behram thought this way about
Einstein, since Behram's own career had focused on general relativity
ever since Behram's graduate student days at the University of Cambridge
in the early 1950s, with general relativity being at the very center of
Behram's research during his entire professional career. Behram's own
particular interest had been on a geometric unification of gravitation
first with electromagnetism and subsequently with all of the other
fundamental forces, an issue to which Einstein himself devoted much of
his own career; and to establish the context of Behram's own work, it is
useful to recall the main ways in which such a unification by metrication
has historically been sought.

Characteristic of attempts to achieve unification through metrication was
the need to find some generalization of general relativity, akin to the
way general relativity had itself generalized special relativity.
Specifically, while special relativity provided a single comprehensive
framework for treating both electromagnetism and Newton's laws of motion
(by modifying the latter), special relativity did not encompass
gravity, with Newton's law of gravitation not being invariant under
Lorentz transformations. The remedy for this was not to come up with
a gravitational law of force which (like the Lorentz electromagnetic
law of force) would be Lorentz invariant, but to instead enlarge
the set of invariances to be obeyed by nature, with general coordinate
invariance replacing Lorentz invariance, and with the metric
$g_{\mu\nu}(x)$ emerging as the gravitational field. 
 
The first attempt at unification through metrication was developed by
Weyl\cite{Weyl1918} as early as 1918, only two years after Einstein's
development of general relativity. Weyl introduced a local transformation
he referred to as a \lq\lq gauge" transformation under which the metric
and the electromagnetic field would transform as 
\begin{equation}
g_{\mu\nu}(x) \rightarrow e^{2\alpha(x)} g_{\mu\nu}(x)
\label{1}
\end{equation}
\begin{equation}
A_{\mu}(x) \rightarrow A_{\mu}(x)-e\partial_{\mu}\alpha(x)~~,
\label{2}
\end{equation}
with gravitation and electromagnetism thus unifying by sharing a common
$\alpha(x)$. Given such a joint transformation, in addition Weyl departed
from the Riemannian geometry of general relativity and replaced it by a
new geometry, \lq\lq Weyl Geometry", in which the connection was
generalized to the $A_{\mu}$-dependent
\begin{equation}
\hat{\Gamma}_{\mu\nu}^{\lambda}={
1 \over 2}g^{\lambda\sigma}\left(\partial_{\mu}g_{\sigma\nu}
+\partial_{\nu}g_{\sigma\mu}-\partial_{\sigma}g_{\mu\nu}\right)
+{1 \over e}g^{\lambda\sigma}\left(g_{\sigma\nu}A_{\mu}
+g_{\sigma\mu}A_{\nu}-g_{\mu\nu}A_{\sigma}\right)~~,
\label{3}
\end{equation}
so that instead of being zero, the covariant derivative of the metric
would instead be given by 
\begin{equation}
g^{\mu\nu}_{\phantom{\mu\nu};\nu}=\partial_{\nu}g^{\mu\nu}
+\hat{\Gamma}_{\nu\sigma}^{\mu}g^{\sigma\nu}
+\hat{\Gamma}_{\nu\sigma}^{\nu}g^{\sigma\mu}
={2A^{\mu}\over e}~~;
\label{4}
\end{equation}
with Eq. (\ref{4}) having the remarkable property of being invariant under
the joint Eqs. ({\ref{1}) and (\ref{2}) (a transformation under which
$\hat{\Gamma}_{\mu\nu}^{\lambda}$ transforms into itself). With the
covariant derivative of a tensor transforming as a vector, Weyl was thus
able to connect $g^{\mu\nu}$ and $A^{\mu}$ in an intricate geometrical
fashion, though at the price of departing from Riemannian geometry. As a
theory Weyl's theory has two noteworthy aspects. First, it introduced the
notion of a gauge transformation, with its usage in Eq. (\ref{1})
entailing a change in the magnitude of $g^{\mu\nu}$ and thus of its size
or gauge. (In the parlance of modern gauge theories, it is a change in
the (complex) phase of a field which is now known as a gauge
transformation, with Eq. (\ref{2}) continuing to be known as a gauge
transformation, but with Eq. (\ref{1}) now being referred to as a scale
transformation.) Second, if imposed as a symmetry, such a scale
invariance would require all mass parameters to be zero identically, and
thus exclude any fundamental Newton constant $G_N$. With it thus
possessing neither the Einstein equations or Riemannian geometry, Weyl's
theory was not so much a generalization of Einstein gravity, but rather a
fairly substantial departure from it, and has not been followed in the
literature.\footnote{There is however, a less radical version of Weyl's
theory in which the invariance of Eq. (\ref{1}) is retained in a geometry
which is a strictly Riemannian one in which
$g^{\mu\nu}_{\phantom{\mu\nu};\nu}=0$. Such a theory is known as
conformal gravity and has been pursued by many authors as a candidate
gravitational theory, one in which mass scales and $G_N$ are induced
dynamically. Such a theory has actually been one of my own research
interests, one which Behram had graciously invited me
to report on at several Coral Gables conferences, with the theory
being found capable of readily addressing the dark matter and dark energy
problems which currently challenge the standard Newton-Einstein
gravitational theory.}

The second attempt at unification through metrication is the now very
familiar work of Kaluza and Klein based on a role for spacetime dimensions
beyond four, a program which later evolved into superstring theory. In the
embryonic work of Kaluza and Klein themselves spacetime was envisaged as
being 5-dimensional, with its thus 15-dimensional gravitational field
decomposing into a 10-component $g_{\mu\nu}$ which was to serve as the
standard 4-dimensional spacetime gravitational field, a 4-component
$g_{5\mu}=A_{\mu}$ to serve as the standard electromagnetic field, thus
leaving over a 1-component $g_{55}=\phi$ which would transform as a
4-dimensional spacetime scalar, a field which is now identified as a Higgs
field. With such Kaluza-Klein theories being based on the 5-dimensional
Einstein equations, they immediately possessed a fundamental $G_N$,
albeit one which was (unlike the role played by $e$ in Eq. (\ref{4})
above) not connected to any of the intrinsic structure of the theory.
However, the theory did have one nice feature in regard to mass scales,
namely the compactification of the fifth of the five dimensions would lead
to a compactification radius and thus to a dynamical mass scale for
particles; though if the compactification radius is taken to be
given by the Planck length scale $(\hbar G_N/c^3)^{1/2}$ associated with
$G_N$, such extra dimensions would be way too miniscule to be
detectable.\footnote{While quite interested in the Kaluza-Klein theory
because of its unification aspects, it is of interest to note that
Einstein nonetheless cautioned\cite{Einstein1931} that \lq\lq Among the
considerations which question this theory stands in the first place: It
is anomalous to replace the four-dimensional continuum by a
five-dimensional one and then subsequently to tie up artificially one of
these five dimensions in order to account for the fact that it does not
manifest itself". It is thus of contemporary interest to note that
with the recent advent of the large extra dimension, brane-localized
gravity program of Randall and Sundrum\cite{Randall1999} this concern of
Einstein may finally have been addressed.} 

The third attempt at unification through metrication is the
non-symmetric gravity program followed by Behram
himself\cite{Kursunoglu1952,Kursunoglu1991},\footnote{While Behram
concentrated on non-symmetric unification throughout his entire career,
at the 2003 Coral Gables conference Dr. John Brandenburg reported on some
recent work done in conjunction with Behram on unification through higher
dimensions.} an approach also seriously considered at one point by both
Einstein\cite{Einstein1925,Einstein1950} and
Schr\"odinger\cite{Schrodinger1950}, and also worked on more recently by
Professor John Moffat an attendee at the 2003 Coral Gables
conference.\footnote{The paper of Moffat and Boal\cite{Moffat1975}
provides reference to Behram's early work on the subject.} On noting that
without symmetrization, a general 4-dimensional spacetime rank two tensor
$\hat{g}_{\mu\nu}$ would contain 10 symmetric and 6 anti-symmetric
components, we see that in the 16 components of
$\hat{g}_{\mu\nu}$ there is precisely the number of degrees required for
a 10-component symmetric gravitational $g_{\mu\nu}$ and a 6-component
anti-symmetric electromagnetic $F_{\mu\nu}$. Unification of gravity with
electromagnetism can thus be achieved by working in a geometry in which
the full $\hat{g}_{\mu\nu}$ metric is not required to be symmetric. To
decompose the full $\hat{g}_{\mu\nu}$ into its symmetric and
anti-symmetric pieces one can thus set 
\begin{equation}
\hat{g}^{\mu\nu}=g^{\mu\nu}+{F^{\mu\nu}\over q}
\label{5}
\end{equation}
where $q$ is an appropriate parameter. What makes this decomposition
so interesting is that with $g_{\mu\nu}$ being dimensionless,
the parameter $q$ has to have the same dimension as the electromagnetic
field strength, viz. $({\rm mass/length/time^2})^{1/2}$. Consequently, and
unlike Weyl's approach, a unification based an a non-symmetric
$\hat{g}_{\mu\nu}$ requires the presence of an intrinsic dimensionful
parameter from the outset; and on setting 
\begin{equation}
q={c^2 \over r_0(2G_N)^{1/2}}~~,
\label{6}
\end{equation}
we see that there has to be an intrinsic length scale, $r_0$, in the
theory, one defined above purely in terms of classical quantities.
Moreover, in a unification which is to include quantum mechanics as
well, this length scale could be identified as the fundamental Planck
scale via
\begin{equation}
r_0=\left({\hbar G_N \over c^3}\right)^{1/2}~~,
\label{7}
\end{equation}
though in his work Behram considered other options for the value of
$r_0$, even including the pure imaginary ones allowed of $r_0$ and $q$
since their overall phases were only fixed by the requirement that the
product $r_0^2q^2$ be identified as the real and positive $c^4/2G_N$. In a
unification based on Eq. (\ref{5}) then, not only do we unify gravitation
with electromagnetism, we additionally see an intrinsic and indispensable
role for Newton's constant from the very outset.

To see how things work in the theory it is convenient to first consider a
theory with just the regular symmetric $g_{\mu\nu}$ and Lagrangian
\begin{equation}
L=g^{1/2}\left[g^{\mu\nu}R_{\mu\nu}+{G_N \over
c^4}\phi^{\mu\nu}(\phi_{\mu\nu}-2F_{\mu\nu})\right]~~,
\label{8}
\end{equation}
where $F_{\mu\nu}$ is the usual electromagnetic tensor and
$\phi_{\mu\nu}$ is an auxiliary antisymmetric field which is not writable
as the covariant curl of a vector. Euler-Lagrange variation of this
theory yields
\begin{eqnarray}
\partial_{\mu}(g^{1/2}\phi^{\mu\nu})&&=0~~,~~F_{\mu\nu}=\phi_{\mu\nu}~~,~~
\partial_{\sigma}\phi_{\mu\nu}
+\partial_{\nu}\phi_{\sigma\mu}
+\partial_{\mu}\phi_{\nu\sigma}=0~~,
\nonumber \\
R_{\mu\nu}-{1 \over 2}g_{\mu\nu}R&&={2G_N \over
c^4}T_{\mu\nu}~~,~~T_{\mu\nu}={1
\over 4}g_{\mu\nu}\phi^{\alpha\beta}\phi_{\alpha\beta}-
\phi_{\mu\alpha}\phi^{\alpha}_{\phantom{\alpha}\nu}~~,
\label{9}
\end{eqnarray}
to thus yield as solution the standard Einstein-Maxwell equations of
motion. Now because both $g^{\mu\nu}F_{\mu\nu}$ and
$\phi^{\mu\nu}R_{\mu\nu}$ are kinematically zero, the Lagrangian of Eq.
(\ref{9}) can be replaced by the equivalent Lagrangian
\begin{equation}
L=g^{1/2}\left[\left(g^{\mu\nu}+{1 \over
q}\phi^{\mu\nu}\right)\left(R_{\mu\nu}-{1 \over
r_0^2q}F_{\mu\nu}\right)+{1
\over 2r_0^2q^2}\phi^{\mu\nu}\phi_{\mu\nu}\right]~~,
\label{10}
\end{equation}
with the replacing of $g^{\mu\nu}$ by $g^{\mu\nu}+\phi^{\mu\nu}/q$
thus still allowing one to recover the standard
Einstein-Maxwell equations of motion in an ordinary $g^{\mu\nu}$
symmetric geometry.

While the above analysis shows that it is permissible to replace
$g^{\mu\nu}$ by $g^{\mu\nu}+\phi^{\mu\nu}/q$, the resulting equations of
motion do not put electromagnetism and gravity on a completely equivalent
footing, since even while the Maxwell tensor serves as the source of the
Einstein tensor, gravity does not serve as a source for electromagnetism.
To remedy this Behram departed from ordinary Einstein gravity with its
symmetric $g_{\mu\nu}$, and instead went to a 16-component
$\hat{g}_{\mu\nu}$, a theory in which tensor manipulations would then
have to be defined anew, in much the same manner as Weyl had to define
the $A_{\mu}$-dependent $\hat{\Gamma}_{\mu\nu}^{\lambda}$ of Eq. (\ref{3})
in his theory. To facilitate the construction of the needed geometry,
Behram went beyond the standard $e^{\mu a}$ vierbein formulation of
gravity introduced by Weyl in 1928 (in which $g^{\mu\nu}=\eta_{ab}
e^{\mu a}e^{\nu b}$) by introducing a second vierbein field
$f^{\mu a}$. In terms of this second vierbein the electromagnetic tensor
can then be introduced as
\begin{equation}
F^{\mu\nu}=\eta_{ab}\left(e^{\mu a}f^{\nu b}-e^{\nu a}f^{\mu b}\right)~~,
\label{11}
\end{equation}
while the contravariant $\hat{g}^{\mu\nu}$ can be introduced as 
\begin{equation}
\hat{g}^{\mu\nu}={1 \over \surd 2}\left[\eta_{ab}e^{\mu a}e^{\nu b}
+{1 \over q^2}\eta_{ab}f^{\mu a}f^{\nu b}
+{1 \over q}\eta_{ab}\left(e^{\mu a}f^{\nu b}-e^{\nu a}f^{\mu
b}\right)\right]~~,
\label{12}
\end{equation}
with its covariant counterpart having symmetric and anti-symmetric
parts which are once and for all defined as
\begin{equation}
\hat{g}_{\mu\nu}=g_{\mu\nu}+{1 \over q}\phi_{\mu\nu}~~.
\label{13}
\end{equation}
The two vierbeins thus define the theory. As well as serve as the
building blocks of the theory, these two vierbeins additionally provide
the theory with a road to symmetry, since the $SO(3,1)$ invariance of
$g^{\mu\nu}$ to Lorentz transformations in the $e^{\mu a}$ space is not
only augmented by an $S0(2)$ invariance of the quantity on the right-hand
side of Eq. (\ref{11}), the combining of the two vierbeins into one
8-component vector allows it to serve as the fundamental representation
of an 8-dimensional symplectic group. In this way Behram had thus
developed a starting point for an incorporation of the symmetries
of elementary particle physics (the use of complex $r_0$ and $q$ even
allows for a connection to unitary groups), an issue he pursued
relentlessly throughout his life.

In Behram's generalized theory there is a generalized connection
$\Gamma_{\mu\nu}^{\lambda}$, one not symmetric in its two covariant
indices, and it is with respect to this connection that
$\hat{g}_{\mu\nu;\rho}$ is to vanish according to 
\begin{equation}
\hat{g}_{\mu\nu;\rho}=\partial_{\rho}\hat{g}_{\mu\nu}
-\Gamma_{\mu\rho }^{\lambda}\hat{g}_{\lambda\nu}
-\Gamma_{\rho \nu}^{\lambda}\hat{g}_{\mu \lambda}=0~~,
\label{14}
\end{equation}
and it is with respect to this connection that the Riemann tensor is to
be constructed as 
\begin{equation}
\hat{R}^{\sigma}_{\phantom{\sigma}\mu\nu\rho}
=-\partial_{\rho}\Gamma_{\mu\nu}^{\sigma}
+\partial_{\nu}\Gamma_{\mu\rho}^{\sigma} -\Gamma_{\mu
\nu}^{\lambda}\Gamma_{\lambda\rho}^{\sigma} +\Gamma_{\mu
\rho}^{\lambda}\Gamma_{\lambda\nu}^{\sigma}~~.
\label{15}
\end{equation}
For such a Riemann tensor the Ricci tensor $\hat{R}_{\mu\nu}$ would
possess both a symmetric ($R_{\mu\nu}$) and an anti-symmetric
($\bar{R}_{\mu\nu}$) piece. Given this structure Behram then introduced a
generalized Lagrangian ($\hat{g}={\rm det}\hat{g}_{\mu\nu}$, $g={\rm
det}g_{\mu\nu}$)
\begin{equation}
L=\hat{g}^{1/2}\hat{g}^{\mu\nu}\left(\hat{R}_{\mu\nu}-{1 \over
r_0^2q}F_{\mu\nu}\right)+{2 \over
r_0^2}\left(\hat{g}^{1/2}-g^{1/2}\right)~~.
\label{16}
\end{equation}
Its Euler-Lagrange variation leads to the same Einstein equation
for the symmetric $R_{\mu\nu}$ as given before in Eq. (\ref{9}), but
with the
$F_{\mu\nu}=\phi_{\mu\nu}$ and
$\partial_{\nu}(g^{1/2}\phi^{\mu\nu})=0$ equations
being replaced by 
\begin{eqnarray}
\phi_{\mu\nu}&&=F_{\mu\nu}-r_0^2\bar{R}_{\mu\nu}~~,
\nonumber \\ 
\left[g^{1/2}\phi^{\nu}_{\phantom{\nu}\mu}\right]_{||\nu}&&=0~~,
\label{17}
\end{eqnarray}
where the $||$ symbol denotes a very particular geometric
derivative\cite{Kursunoglu1991} that Behram was able to construct in his
theory. As such Eq. (\ref{17}) thus provides the intricate connection
between electromagnetism and geometry which Behram had
sought.\footnote{As noted in Pais's book on Einstein\cite{Pais1982} (a
book which also references Behram's work), one of Einstein's motivations
for turning to non-symmetric unified theories was to try to find a theory
whose particle-like solutions would be singularity-free. Interestingly,
Behram was able to show\cite{Kursunoglu1991} that the solutions to
the theory based on Eq. (\ref{16}) were singularity-free.}

As well as work on unification through metrication, Behram will also be
remembered for the Coral Gables Conference Series which have been
occurring over a span of almost forty years. The very first of
these conferences took place in 1964\footnote{This was an auspicious year
for me personally as it was my first year in graduate school at the
Weizmann Institute, and even now I can still recall the excitement of
that period when Professors Yuval Ne'eman, Harry Lipkin and (frequent
Weizmann visitor and Coral Gables Conference Series co-founder) Sydney
Meshkov would report on the Coral Gables conferences.} at which  Behram
presented a paper \lq\lq A new symmetry group for elementary particles".
As a conference organizer Behram was indefatigable in continuing this
conference series over the years, while also never flagging as a
researcher. Indeed, he continued to make original presentations at the
Coral Gables conferences right up to the end.  Thus he presented \lq \lq
Recent developments in gravitational theory and an intrinsic cosmological
parameter" at the 2000 Coral Gables conference,\footnote{It is of
interest to note that the article which appeared immediately following
Behram's paper in the 2000 proceedings was given by none other than the
sixtieth birthday honoree of the 2003 conference Professor Paul Frampton
who in 2000 had presented a paper \lq\lq Quintessence and cosmic
microwave background". As usual, here was Paul working creatively on one
of the hot topics of the day, and I personally wish him many productive
years to come.} \lq \lq New Facts on the nature of gravitational force and
non-linear oscillations of space" at the 2001 Coral Gables
conference,\footnote{It is of interest to note that the article which
appeared immediately following Behram's paper in the 2001 proceedings was
given by Behram's long-time associate and Coral Gables Conference Series
co-founder Professor Arnold Perlmutter who had presented an application
of Behram's theory: \lq\lq A conjecture on the existence of attractive
and repulsive gravitational forces in the generalized theory of
gravitation". There is perhaps no higher compliment that one can pay a
colleague than the one Arnold thus gave to Behram by working on his
ideas.} and \lq \lq The layered structure of the universe" at the 2002
Coral Gables conference. 

Beyond physics, Behram also established and ran the Global Foundation
which organized meetings on and concerned itself with the great
social and international issues of the day. He was an urbane and cultured
human being, one who can truly be called a Renaissance man. During his
lifetime Behram championed many causes, but the one which I suspect may
have given him the most pleasure was the eventual coalescence of
elementary particle physics and gravitation, something he presciently
first advocated more than fifty years ago, and patiently continued to
persistently champion ever thereafter.

\end{document}